\documentclass[rmp,twocolumn,nofootinbib,amsmath,amssymb]{revtex4}

\usepackage{aas_macros}

\begin{document}
\title{Addressing LISA Science Analysis Challenges}
\author{Matthew J. Benacquista}
\email{benacquista@msubillings.edu}
\affiliation{Montana State University, Billings, MT 59101 USA}
\author{Lee Samuel Finn}
\email{LSFinn@PSU.Edu}
\affiliation{Center for Gravitational Wave Physics, 
The Pennsylvania State University, University Park, PA 16802, USA}
\author{Shane L. Larson}
\email{shane@gravity.psu.edu}
\author{Louis J. Rubbo}
\email{rubbo@gravity.psu.edu}
\affiliation{Center for Gravitational Wave Physics, 
The Pennsylvania State University, University Park, PA 16802, USA}
\date{\today}

\begin{abstract}
    
\end{abstract}
\maketitle


\section{Introduction}\label{sec:introduction}

\emph{Science analysis} is the process by which observations 
are transformed into scientific insight and understanding. In the 
same way that even the most artful analysis cannot compensate 
for poor data, so even the best instruments and observational 
skill can compensate for an inability to adequately analyze the 
data. LISA is no exception. 

Exactly because LISA is a pathfinder for a new scientific discipline 
--- gravitational wave astronomy --- LISA data processing and 
science analysis methodologies are in their infancy and require 
considerable maturation if they are to be ready to take advantage 
of LISA data. Here we offer some thoughts, in anticipation of the 
LISA Science Analysis Workshop, on analysis research problems
that demonstrate the capabilities of different proposed analysis 
methodologies and, simultaneously, help to push those techniques
toward greater maturity. Particular emphasis is placed on
formulating questions that can be turned into well-posed 
problems involving tests run on specific data sets, which can 
be shared among different groups to enable the comparison 
of techniques on a well-defined platform. 

The questions, from which demonstration problems can be posed, 
are organized by source type. Accompanying each set of questions 
is a short discussion meant to provide context and motivation for the 
questions that follow.

\section{Technology Readiness Levels}
One way to measure the maturity of LISA data processing and 
science analysis technology techniques 
is to use the NASA Technology Readiness Level (TRL) metric.  TRLs
provide a systematic measurement of the maturity of a particular
technology (hardware or software) relative to mission
goals~\cite{Mankins:1995:trl}.  Table \ref{tbl:trl} describes the NASA
TRLs for software.  When LISA science data becomes available the
software necessary for data processing and science analysis related to
LISA science requirements should be at least TRL~7 and preferably 
TRL~8.  When LISA science results are released the software should 
be at TRL 8.  

We are aware of no LISA analysis methodologies beyond
TRL 2 and \emph{the principal goal of the questions posed here is to 
point the way toward elevating the TRL level of LISA analysis technology.} 
For these questions to be useful in this regard they must be attuned to
the present level of analysis sophistication. Thus, the problems
described here are focused on demonstrating capability at the level of
TRL 2 or TRL 3. Later demonstration problems will focus on further 
developing data processing and science analysis technologies to higher 
TRLs.

\begin{table*}
\caption{NASA Technology Readiness Levels for software.}\label{tbl:trl}
\begin{tabular}{|p{0.5in}|p{4.75in}|}
\hline
TRL 1&Basic principles observed and reported. \emph{Basic properties
  of algorithms, representations \& concepts. Mathematical
  formulations. Mix of basic and applied research.}\\
\hline
TRL 2&Technology concept and/or application formulated. \emph{Basic
  principles coded. Experiments with synthetic data. Mostly applied
  research.}\\
\hline
TRL 3&Analytical and experimental critical function and/or
characteristic proof-of-concept. \emph{Limited functionality
  implementations. Experiments with small representative data
  sets. \underline{Scientific feasibility fully demonstrated.}}\\
\hline
TRL 4&Module and/or subsystem validation in laboratory
environment. \emph{Standalone prototype implementations.  Experiments
  with full-scale problems or data sets.}\\
\hline
TRL 5&Module and/or subsystem validation in relevant
environment. \emph{Prototype implementations conform to target
  environment/interfaces. Experiments with realistic
  problems. Simulated interfaces to existing systems.}\\
\hline
TRL 6&System/subsystem prototype demonstration in a relevant
end-to-end environment. \emph{Prototype implementations if the
  software is on full-scale realistic problems.  Partially integrated
  with existing hardware/software systems.  Limited documentation
  available.  \underline{Engineering feasibility fully
    demonstrated.}}\\
\hline
TRL 7&System prototype demonstration in high-fidelity environment
(parallel or shadow mode operation). \emph{Most of the software is
  functionality available for demonstration and test.  Well integrated
  with operational hardware/software systems.  Most software bugs
  removed.  Limited documentation available.}\\
\hline
TRL 8& Actual system completed and ``mission qualified'' through test
and demonstration in an operational environment. \emph{Thoroughly
  debugged software.  Fully integrated with operational hardware and
  software systems.  Most user documentation, training documentation,
  and maintenance documentation completed.  All functionality tested
  in simulated and operational scenarios.  Validation \& Verification
  completed.}\\
\hline
TRL 9&Actual system ``mission proven'' through successful mission
operations.  \emph{Thoroughly debugged software.  Fully integrated
  with operational hardware/software systems.  All documentation has
  been completed and users have successful operational experience.
  Sustaining software-engineering support in place.
  \underline{Actual system fully demonstrated.}}\\
\hline
\end{tabular}
\end{table*}

\section{Verification Binaries}\label{sec:verification}

The verification binaries are a unique subset of the resolved galactic
binaries described in the next section.  Verification binaries are systems
that have been identified pre-science operation and that are
well characterized through more traditional astronomical observations.
This characterization of the verification binaries makes it possible 
make it possible to accurately \emph{predict} the strength, polarization, 
and propagation direction of the gravitational waves from the source. 
LISA's response and function can thus be verified from its observations 
of these systems. 

The verification sources will be among the first targets in a search
of the LISA data.  The results of those searches will be used to
validate and confirm the performance and expectations for the
software, instrumental noise, and hardware performance of the
observatory.  As such, these binaries will play a vital role in
characterizing early LISA performance, and specific analysis will need
to be developed to address this special population of sources.
Questions of particular interest include:

\begin{itemize}
\item How soon after observations begin can you identify a verification
binary, ignoring other sources? With other sources (binaries, 
supermassive black holes, extreme mass ratio inspirals, etc)?

\item How does knowledge of a verification binary's parameters
change as a function of LISA observing time?  How long must LISA
observations last to recover the verification parameters to the level of
accuracy provided by electromagnetic observations? 
\end{itemize}

Early studies on LISA observations of verification binaries have
started~\cite{Stroeer:2006:lvb}.  Prospective LISA verification
binaries have been identified and a database of the current known
parameters for these binaries is being 
maintained~\cite{Nelemans:2006:wiki} for use by the LISA community.

\section{Galactic Binaries}\label{sec:gxBinaries}

Stellar mass galactic binary systems are the most abundant of the
sources LISA is capable of observing. Crude estimates place the number
of binaries that LISA can resolve as distinct sources in the tens of 
thousands \cite{Nelemans:2001:gws, Timpano:2005:cgg}, with millions
more forming an unresolvable background at lower frequencies. The
large population of resolvable binaries provides opportunities to
develop a more complete map of the galaxy, study the mass distribution
of binary components, and study the population and evolution of mass
transfer systems.  The unresolvable binary background provides
additional information about the number of binaries and their galactic
distribution. Finally, because the signal from binaries is
ever-present, signals from other sources must be identified and
characterized in the forest of resolvable binaries and the fog of the
confusion background. The ability to identify and characterize
isolated binaries and the confusion background is thus a crucial first
challenge for LISA science analysis.

\subsection{Isolated binaries}
In addition to the verification binaries described in $\S$
\ref{sec:verification}, there will be several thousand resolvable
binaries which will be unknown and uncharacterized before LISA begins
observations.  The ability to identify, characterize, and extract
science from observations of these binaries will depend largely on the
analysis technique used.  Specific questions which are of interest for
an individual binary analysis algorithm are:
\begin{itemize}
\item Given a realistic galactic model, how many individual binary sources 
can be resolved?  How accurately can resolved binaries be characterized?  
How does the characterization change as observing time increases?  Does 
the method mistakenly identify ``false binaries''?

\item How accurately can the different binary parameters ($\sin i$,
amplitude, etc)  be determined as function of SNR, sky location and
observing time?

\item Given a particular analysis technique, what is LISA's
``resolving power''?  How well can the technique spatially resolve
individual binaries on the sky?  How well can individual binaries be
resolved in frequency?

\end{itemize}

A number of analysis techniques targeting isolated binaries have
appeared in the literature~\cite{Krolak:2004:rsl, Cornish:2003:lda,
Umstatter:2005:bms}.  These techniques have explored a variety of
approaches with regard to identification and parameterization of
binaries; they have yet to be compared and contrasted directly.

\subsection{Confusion background}
Below some frequency every analysis techniques targeting 
individual binary sources will break down as overlapping signals 
from the millions of short period binaries in the galaxy merge to 
form a \emph{confusion-limited background}.  

The confusion-limited background is both a boon and a bane. 
The background amplitude, shape, and angular distribution 
depends on the astrophysics of binary evolution, the total number 
of binaries contributing to the confusion, and the shape of the galaxy.  
By measuring this background amplitude, spectrum and angular 
distribution on the sky we are measuring these characteristics of
our galaxy. On the other hand, the confusion-limited background 
is an \emph{astrophysical} source of noise that limits our ability to
identify other sources at low frequencies. Understanding the 
onset of confusion will play an important in understanding the 
low-frequency science that is possible with LISA observations. 
Interesting questions that can be posed of techniques targeting 
the confusion limit include:
\begin{itemize}
\item How well can the spectrum (shape and level) of the confusion
noise be determined as a function of frequency and the 
confusion spectral density?

\item How well can the spatial distribution of the confusion be
determined? 

\item How does the characterization of the confusion spectrum evolve
with increased observing time?
\end{itemize}

A great deal of \emph{astrophysical} analysis has gone into \emph{predicting}
the possible populations that will contribute to the confusion limited
background. A variety of techniques have been considered to begin to
approach the question of \emph{how} LISA will view the
background~\cite{Seto:2004:amg, Edlund:2005:wdw, Mohanty:2006:tar}.

\subsection{From isolated to confused}
The number density of galactic binaries increases rapidly with decreasing
frequency; thus, at high frequencies we have isolated binaries while at 
low frequencies the binaries are unresolvable and we will not be able to 
identify the signal from a single binary. How the fraction of resolvable 
binaries decreases with decreasing frequency directly affects our ability 
to observe sources that may be situated in the transition band. 
\begin{itemize}
\item Given a realistic model of the galactic binary distribution, how
does confusion ``emerge'' as a function of frequency (binary period)? 

\item How does the ``fog'' of confusion ``lift'' as LISA observations 
progress?

\item There will always be exceptionally bright sources, which stand-out
above the confusion. How does the number of such exceptional binaries 
vary with frequency? 
\end{itemize}

An important element in research studies that target problems
relating to the galactic binaries is the availability of galactic
realizations.  Several different realizations exist, such as those
built from binary distribution functions \cite{Hils:1990:grg,
Timpano:2005:cgg}, and those derived from population synthesis models
\cite{Benacquista:2004:sli, Nelemans:2001:gws, Edlund:2005:wdw}.

\section{Burst Signals of Astrophysical Origin}
LISA can be expected to observe bursts of gravitational
waves from relativistic fly-bys of compact objects about supermassive black
holes~\cite{Rubbo:2006:ere}. More speculative is the radiation from 
the disruption of a main sequence or white dwarf via a too-close encounter with 
an intermediate mass black hole. Still more speculative is radiation 
from topological defects in cosmic strings~\cite{Damour:2000:gwb, 
Damour:2001:gwb, Damour:2005:grc}. Specific questions of interest for
analysis methods that target burst gravitational wave sources include
\begin{itemize}
\item How well are individual bursts resolved in the LISA data as a
  function of signal-to-noise and burst duration?

\item Is it possible to distinguish a noise burst in the measurement or
sensing functions of the constellation from a burst arising from an 
astrophysical source? 

\item Can burst sources of radiation be characterized well enough 
that they can be distinguished by source or source type? 
\end{itemize}

\section{Extreme Mass Ratio Inspirals}

When studying spacetimes, it is natural to discuss the motion of a
test particle in the background spacetime of interest.  Nature has
been kind enough to provide systems that strongly approximate
the test body case in the extreme mass ratio inspirals (EMRIs): the
capture of a stellar mass compact secondary object by an intermediate
or supermassive black hole. With each orbit gravitational waves
carry away energy and angular momentum and, at least while 
the rate of loss of energy and angular momentum is small, the 
secondary can be thought to evolve along a trajectory of geodesics. 
By studying this evolution it may be possible to reconstruct a broad
family of geodesics and thus ``map'' the spacetime in the neighborhood
of a black hole~\cite{Ryan:1997:aem}.

EMRI radiation is not necessarily continuously observable in the LISA band. 
When the orbits are relativistic the radiation is beamed, leading to large
amplitude variations as the beam follows the secondary in its orbit. Additionally, 
many EMRIs may be in high eccentricity orbits, in which case the radiation 
may only be in the LISA band during a small fraction of the orbit. 

Besides being natural laboratories for conducting tests of general
relativity, the event rate and characteristics of EMRIs can lead to
insights to the structure and evolution of galactic centers. EMRIs
allow high precision estimates for the central black hole's mass and
spin~\cite{Barack:2004:lcs}.  The event rate alone gives an indication
of the stellar density in the cores of galaxies.

The apparent difficulty associated with detecting EMRIs is that each
system is parameterized by up to fourteen parameters.  The high
dimensionality of the parameter space hinders the blunt use of
standard template matching techniques.  Consequently, alternative
approaches to the EMRI detection and characterization problems are
required.  Early analysis methods have included semi-coherent
searches~\cite{Gair:2004:ere}, and the use of time-frequency
methods~\cite{Wen:2005:dem, Gair:2005:dem}.  These approaches are
promising, but are still in the early formulation stages.

Central issues in EMRI data analysis are:
\begin{itemize}
\item For EMRIs that lead to periodic bursts of radiation in the LISA band
(owing either to orbital eccentricity or beaming) can multiple bursts from 
a single EMRI system be linked with each other?
\item What features of an EMRI signal (i.e. location, black hole spin,
  secondary mass, etc.) become ``in focus'' with increased waveform
  model complexity, signal-to-noise ratio, and/or observation duration?
\item How well can a complete EMRI signal be identified and
  characterized in the presence of instrument noise? A confusion-limited
  background? A confusion-limited background of EMRIs? 
\end{itemize}

\section{Massive Black Hole Binaries}
Observing the inspiral, coalescence and ringdown of massive black hole
binaries will provide critical clues to the order in which the large
scale structure in the Universe evolved: did stars evolve and then
galaxies, or galaxies and then stars?  Did supermassive black holes
form hierarchically from run-away collision of lower-mass black holes,
or were they massive at birth, forming from the collapse of primordial
clouds of gas? LISA can help answer these questions by producing 
a census of merger events mass and luminosity distances. To obtain 
luminosity distances it will be necessary to have accurate sky positions. 
For gravitationally ``bright'' sources this may come from the gravitational
wave observations themselves \cite{Schutz:1986:DHC,Finn:1996:BIG}; 
however, for dimmer sources the gravitational wave estimates of position 
may be too crude for an accurate distance determination, in which case 
the observation of an optical counterpart (i.e., the galaxy host of the merger)
will be necessary to get an accurate redshift \cite{Kocsis:2006:fec}. 

While the inspiral, coalescence and ringdown of a supermassive black
hole will always be detected in the presence of the galactic binaries,
if we can't identify and characterize a MBH binary source all by
itself we'll never be able to identify and characterize a MBH binary
in the presence of the galactic binary forest and confusion
background. Therefore, each of the following three questions should be
answered at three levels: (1) in the absence of a galactic binary
confusion background, (2) in the presence of an artificially
``cleaned'' background with all bright sources removed, and (3) in the
presence of a full galactic binary background:
\begin{itemize}
\item How well can an SMBH binary be identified and characterized?

\item How ``bright'' must a MBH binary be to be identified? How does
the accuracy of the MBH characterization scale with ``brightness''?

\item How well, as a function of observation time, can you determine
where and when the binary will coalesce? (i.e., what precision a month
from coalescence? a week? a day?)
\end{itemize}


\section{Multi-source challenges}

The identification of every LISA source will take place in the 
simultaneous presence, in the LISA data, of millions of long-period
galactic binaries, myriads of distinctly resolvable short-period
galactic binaries, and multiple extreme-mass-ratio inspirals and
supermassive black hole inspirals.  A critical challenge for LISA
analysis is the ability to identify and characterize these sources in
each others presence. Central questions in multi-source analysis include
\begin{itemize}
    \item How well can different source types in the data be searched for
    \emph{sequentially}?  For example, can SMBH binaries be found and
    subtracted out of the data before galactic binaries or EMRIs are
    searched for?
    
    \item How well can different source types in the data be searched for
    \emph{simultaneously}?
    
    \item What fidelity is required of theoretical source models
    for a given multi-source science analysis procedure to work?  How 
    does the effectiveness of the analysis method scale with source
    simulation fidelity?
    
    \item How can \emph{source catalogs} of known sources in the LISA 
    data be used to best effect in multi-source analysis?
    
    \item Can source catalogs be created from electromagnetic
    observations in advance of LISA? Can source catalogs be created
    directly from the LISA data?  If they can, how do they change and
    evolve with extended LISA observing periods?
    
    \item How will unmodeled sources be handled by multi-source search
    and characterization procedures?
\end{itemize}

In many ways, multi-source analysis synthesizes all the challenges
related to given source types into a single problem.  This synthesis
represents the inexorable march toward more realistic simulations of
what actual LISA science analysis will look like.

Several groups have started to make forays into analysis of data
segments with strongly overlapping sources, using a variety of modern
algorithms~\cite{Cornish:2005:lda, Crowder:2006:lda, Wickham:2006:mcm,
Cornish:2006:nwc}.

\section{Data sets for science analysis challenges}
Science analysis demonstrations and feasibility studies
require the use of simulated data that is well-characterized and 
of sufficient fidelity that the feasibility demonstration is meaningful. 
Trade studies or evaluations and qualification of different 
technologies are best performed under identical conditions; so, 
there is great value in archiving and sharing data sets used for 
different studies so that different analysis methods
can be characterized under the same conditions and their results
compared. An additional advantage of shared data sets for 
science analysis demonstrations and feasibility studies is
that comparison among studies carried out on the same data but
with different techniques provides practice for the day when 
real LISA data will be available and there is only \emph{one} LISA
data set and all studies will take place on the same data. 

Every demonstration or feasibility study has a goal that determines 
the appropriate degree of fidelity (in noise characteristics, LISA 
simulation approximations, etc) that the simulated data set must
satisfy. The fidelity of the data used in a study should not substantially
exceed that required for a meaningful demonstration in order to avoid 
complications in the study's interpretation. So, for instance, data sets
designed to probe the ability of an analysis technique to resolve pairs
of binary star systems need not be of full bandwidth. To be shareable, 
the data sets should also be complete and fully documented. Completeness, 
in this case, means that the data set should contain everything necessary 
to carry-through the analysis: no assumptions about, e.g., the approximations
made in the simulation (rigid adiabatic LISA? second order eccentricity 
orbits? constellation position and phase at the initial epoch?) or in the 
constellation response (what are the observables? low-frequency
approximate response, or exact response?) should need to accompany 
the data sets. 

Data sets that can be used as a common platform for addressing these
challenges are currently being developed, produced, and distributed by
two groups. The \emph{Testbed for LISA Analysis} (TLA) Project, spearheaded 
by the Center for Gravitational Wave Physics, has developed a data container
(the Simulated LISA Data Product, or SLDP), which was developed to meet
the goal of completeness as described above. The 
\emph{Mock LISA Data Challenges} (MLDC) group, organized by the
LISA International Science Team Working Group 1B, has developed
the LISAxml data container that is complete in a different sense: LISAxml
files include a full description of the source content of the data they contain. Both
groups, which share many members in common, provide software for
reading and writing data sets in these two different format; additionally, the 
TLA Project will provide SLDP versions of the simulated data content of LISAxml 
files provided by the MLDC effort. 

Data sets suitable for addressing several of the science analysis
issues presented in this paper, and in the recommendations that 
emerge from the LISA Science Analysis Workshop, will be made 
available as SLDP files through the Testbed for LISA Analysis web site 
\texttt{<http://tla.gravity.psu.edu>}.  The TLA Project invites the 
participation of scientists in all aspects of its work, from developing
software to support collaborative work in LISA science analysis, to
generating and providing sample data sets for analysis studies, to 
contributing to an annotated bibliography of analysis study results, and 
many things in between. For more information on how to become
involved in the TLA Project visit the TLA website at 
\texttt{<http://tla.gravity.psu.edu/getinvolved/>}.

The MLDC effort has developed a systematic series of 
``challenges'', which are available through their collaborative
working wiki hosted at Caltech, 
\texttt{<http://www.tapir.caltech.edu/dokuwiki/>} (click on LISA
Science Team Working Group 1B). The MLDC Group will 
provide data sets suitable for addressing these challenges as 
LISAxml files through Astrogravs at
\texttt{<http://astrogravs.nasa.gov/>}. People interested
in participating in the MLDC effort should visit their working wiki for
contact information.

\section{Final thoughts}
The principal goal of the \emph{LISA Science Analysis Workshop} is to
encourage the development and maturation of science analysis
technology in preparation for LISA science operations. The 
principal outcome of the workshop will be a report, written by the 
workshop participants, that
\begin{itemize}
\item articulates specific demonstrations of analysis capabilities that can 
(and should!) be addressed by the LISA science analysis community in the 
next 1-2 years; 
\item defines the specific data sets needed to make these demonstrations; 
\item identifies the support structure (software tools, community forums and 
meetings) that simplify the completion of these studies; and
\item provides a forum for the effective communication and dissemination 
of the results of these studies. 
\end{itemize}

LISA's best advocates are the scientists whose blood, toil, tears and sweat
will carry-out the LISA science program, from technology through analysis
and science interpretation. If you are not already involved in LISA science
analysis we urge you to become involved, by joining one or both of the 
TLA and MLDC projects. 

\acknowledgements
We gratefully acknowledge the support of NASA awards NNG05GF71G and NNG04GD52G, 
The Pennsylvania State University Center for Space Research Programs, and The Center 
for Gravitational Wave Physics. The Center for Gravitational Wave Physics is supported by 
the National Science Foundation under cooperative agreement PHY 01-14375. 


\end{document}